\documentclass[conference]{IEEEtran}
\IEEEoverridecommandlockouts

\usepackage{cite}
\usepackage{amsmath,amssymb,amsfonts}
\usepackage{algorithmic}
\usepackage{balance}
\usepackage{graphicx}
\usepackage{textcomp}
\usepackage{hyperref}
\usepackage{microtype}
\usepackage{booktabs}
\usepackage{tabularx}
\usepackage{multirow}

\newcolumntype{C}{>{\centering\arraybackslash}X}
\newcolumntype{L}{>{\raggedright\arraybackslash}X}
\newcolumntype{R}{>{\raggedleft\arraybackslash}X}
\newcolumntype{P}[1]{>{\raggedright\arraybackslash}p{#1}}
\usepackage{siunitx}
\usepackage{etoolbox}                           %
\newcommand{\ubold}{\fontseries{b}\selectfont}  %
\robustify\ubold                                %
\newcommand{\tablecaptionsep}{\vspace*{0pt}}

\usepackage{xcolor}
\def\BibTeX{{\rm B\kern-.05em{\sc i\kern-.025em b}\kern-.08em
    T\kern-.1667em\lower.7ex\hbox{E}\kern-.125emX}}

\definecolor{hermancolor}{HTML}{FF6600}

\definecolor{mattcolor}{HTML}{0004ff}

\begin{document}
\def\modelname{{MARS6}}

\title{MARS6: A Small and Robust Hierarchical-Codec Text-to-Speech Model\\
\thanks{\IEEEauthorrefmark{2}This author is with {E\&E Engineering, Stellenbosch University, South Africa}. All contributions were made in their capacity as an advisor to \mbox{Camb.ai Inc.}}
}

\author{\IEEEauthorblockN{Matthew Baas\IEEEauthorrefmark{1}, Pieter Scholtz\IEEEauthorrefmark{1}, Arnav Mehta\IEEEauthorrefmark{1}, Elliott Dyson\IEEEauthorrefmark{1}, Akshat Prakash\IEEEauthorrefmark{1}, Herman Kamper\IEEEauthorrefmark{1}\IEEEauthorrefmark{2}}
\IEEEauthorblockA{\IEEEauthorrefmark{1}\textit{Camb.ai}\\
Email: research@camb.ai}
}

\maketitle

\begin{abstract}
Codec-based text-to-speech (TTS) models have shown impressive quality with zero-shot voice cloning abilities.
However, they often struggle with more expressive references or complex text inputs.
We present \modelname{}, a robust encoder-decoder transformer for rapid, expressive TTS.
\modelname{} is built on recent improvements in spoken language modelling. %
Utilizing a hierarchical setup for its decoder, new speech tokens are processed at a rate of only 12~Hz, enabling efficient modelling of long-form text while retaining reconstruction quality.
We combine several recent training and inference techniques to reduce repetitive generation and improve output stability and quality.
This enables
the 70M-parameter \modelname{} to achieve similar performance to models many times larger.
We show this in objective and subjective evaluations, comparing TTS output quality and reference speaker cloning ability.
Project page: {\small \url{https://camb-ai.github.io/mars6-turbo/}}
\end{abstract}


\section{Introduction}

Text-to-speech (TTS) systems have improved many-fold in recent years, showcasing new capabilities in speaker cloning cability and naturalness~\cite{casanova2024xtts_interspeech, chen2024valle2neuralcodec, li2024styletts}.
One promising area in TTS is spoken language models (SLMs)~\cite{wang2023neuralcodeclanguagemodels}, where a neural audio codec converts speech into a sequence of discrete tokens.
Like text language models, SLMs are trained to predict the  next discrete token autoregressively, typically using a transformer-based architecture. 
But most prior SLM-based TTS systems exhibit a key limitation -- they are unstable~\cite{hu2024robust, han2024vallerrobustefficient}.
When the reference audio or text is complex or out-of-domain, SLMs often perform poorly compared other TTS methodologies.

While there have been several methods %
proposed %
to address such limitations, they are typically considered in isolation (e.g.\ repetition aware sampling~\cite{chen2024valle2neuralcodec}), or they drastically increase the runtime (e.g.\ multiple sampling~\cite{melle_meng2024autoregressive,chen2024valle2neuralcodec}).
To this end, we propose \modelname{} -- a 70M parameter SLM for robust, rapid and expressive TTS.
We combine several %
recent techniques,
and propose
some new techniques from 
outside the TTS domain (e.g.\ odds ratio preference optimization~\cite{hong2024orpo} and a new top-$p$ fallback sampling mechanism).
\modelname{} consists of an encoder-decoder transformer, and combines a hierarchical speech codec with a hierarchical decoder architecture to process speech tokens at a rate of 12~Hz.
Together with the aforementioned inference techniques, this makes \modelname{} a highly robust and capable TTS model.
It is also a showcase for a `bag of tricks' that we introduce for SLM-based TTS.

For our experiments, 
we construct a difficult in-the-wild TTS evaluation set 
using the expressive EARS dataset~\cite{richter2024ears}.
We compare \modelname{} against prior diffusion- and autoregressive-based TTS models using objective and subjective evaluations.
\modelname{} performs competitively, even against models many times its size.
When used with voice cloning based on a snippet of reference audio, \modelname{} captures the target speaker identity closely, surpassing prior models in subjective speaker similarity evaluations.
Our main contribution is to demonstrate that we can combine
several recently proposed techniques with new techniques proposed herein
during model design, training, and inference, to stabilize outputs and yield a more robust SLM-based TTS system.
Demo, samples, code, and checkpoints: \mbox{\small \url{https://camb-ai.github.io/mars6-turbo/}}.

\section{Related Work}

Within SLMs, there are broadly three ways to approach speech tokenization.
The first is to tokenize speech using acoustic tokens at a fixed sample rate, as done in EnCodec and DAC~\cite{defossez2022highfi,kumar2024high}.
The second is to mix acoustic and semantic tokens using two different quantizers~\cite{baade24_interspeech}, e.g.\ using clustered HuBERT features for semantic and EnCodec for acoustic tokens.
The third, which we explore here, is that of hierarchical acoustic codecs, such as SNAC~\cite{Siuzdak_SNAC_Multi-Scale_Neural_2024}.
These codecs quantize speech into acoustic tokens in different codebooks, each with its own sampling rate.
This makes lower codebooks more `coarse', and higher sample-rate codebooks %
`fine'.
The progenitor SLM TTS model, VALL-E, and its successors~\cite{wang2023neuralcodeclanguagemodels,chen2024valle2neuralcodec,han2024vallerrobustefficient},
uses an autoregressive transformer to predict the most coarse acoustic codebook, and a non-autoregressive model to predict the remaining codebook values. 

Despite success, VALL-E and its descendants often suffer from stability issues.
Several studies have tried to address this~\cite{song2024ellavstableneuralcodec,dang2024livespeech}, e.g.\ by adding linguistic and phonemic constraints to improve coherence between the output speech  and the given input text~\cite{wang2024hamttshierarchicalacousticmodeling}.
But most of these improvements require phoneme alignments during training. 
The `bag-of-tricks' we introduce in this paper does not require such resources.

\section{\modelname{}}

\begin{figure}[t!]
\centering
\centerline{\includegraphics[width=0.99\linewidth]{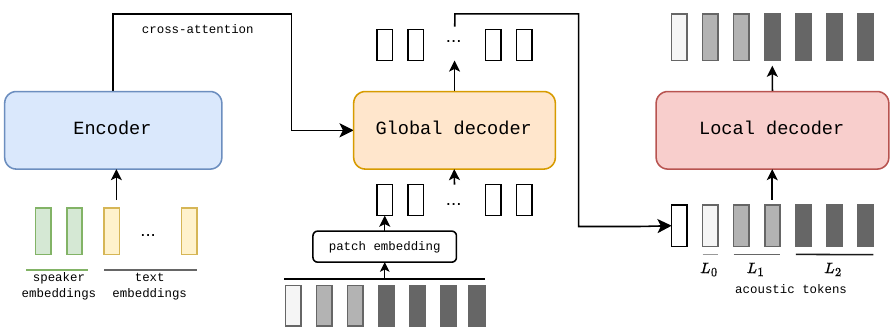}}
\vspace{-2mm}
\caption{
    \modelname{} is an encoder-decoder transformer.
    The encoder converts a speaker embedding and sequence of text embeddings to latent vectors for cross-attention in the global decoder.
    The hierarchical autoregressive decoder has two parts:
    The global decoder produces new latent vectors at a low sample rate, where each vector is autoregressively decoded to acoustic tokens using a smaller local decoder model.
    The entire patch of acoustic tokens then forms the next input vector to the global decoder through a patch embedding. %
}
\label{fig:1_system_diagram}
\end{figure}

\figurename~\ref{fig:1_system_diagram} shows the \modelname{} model, which follows an encoder-decoder architecture.
For zero-shot speaker cloning, the encoder takes in reference speaker embeddings together with the target text.
The decoder is hierarchical and made of two components: a local and global decoder, similar to the proposal of~\cite{yu2024megabyte}.
The global decoder takes input acoustic features in patches, and its output is fed into the local decoder to autoregressively predict all acoustic tokens for the next patch. Details are given next.\footnote{Mars is the Roman god of war. It is also the name of a chocolate bar first produced in 1932. \modelname{} was our sixth internal model version.}

\vspace{-0.9mm}
\subsection{Encoder and input representation} 
\label{subsec:encoder}
\vspace{-0.4mm}

The encoder is a non-causal transformer encoder using Mish activations~\cite{misra2019mish} with sinusoidal positional embeddings, similar to~\cite{vaswani2017}.
Its input sequence consists of two parts.
First, to clone the target speaker, we compute a speaker embedding using a pretrained speaker verification model and a secondary embedding using CLAP~\cite{CLAP2023}.
The former, being trained mostly on non-emotive speech, gives a good base speaker representation.
But, for expressive references where the speaker verifier's embeddings are less meaningful, the more broadly trained (but less speaker-specific) CLAP embedding is useful.
These two vectors are mapped to the dimension of the transformer using a projection layer, and then joined along the sequence length (`speaker embeddings' in Fig.~\ref{fig:1_system_diagram}). 
Second is the sequence of text embeddings corresponding to the desired text being spoken  (`text embeddings' in Fig.~\ref{fig:1_system_diagram}).
To reduce the token count and improve speed, the text is tokenized using byte-pair encoding (BPE)~\cite{gage1994new_bpe}.

To improve reference coherence and output stability, we adapt a lesson from~\cite{allenzhu2024physicslanguagemodels33}.
We give the encoder a way to learn when an output should be high fidelity (e.g.\ 48~kHz audio from VCTK~\cite{Yamagishi2019CSTRVC} downsampled to the 24~kHz codec sampling rate) or lower fidelity (e.g.\ upsampled 16~kHz audiobook data).
To indicate the target quality to the encoder, we prepend the original sample rate to the text, e.g. for 16~kHz, ``Mister \ldots" becomes ``[16000] Mister \ldots ".

\subsection{Global decoder}

\modelname{} operates on hierarchical acoustic tokens from the SNAC acoustic model~\cite{Siuzdak_SNAC_Multi-Scale_Neural_2024}.
SNAC encodes speech into discrete sequences using residual vector quantization with codebooks at different sampling rates, representing different levels in a hierarchy, where earlier codebooks are sampled less frequently.
For \modelname{} we use the 3-codebook SNAC~\cite{Siuzdak_SNAC_Multi-Scale_Neural_2024}, with codebook sample rates of 12 ($L_0$), 24 ($L_1$), and 48~Hz ($L_2$). %

Like the encoder, this decoder uses Mish activations and sinusoidal positional embeddings. 
The global decoder takes patches of acoustic tokens from SNAC at 
12~Hz, whereby all 
codebook tokens generated within  $\frac{1}{12}$s are flattened and fed through a patch embedding~\cite{yu2024megabyte} to yield a 12~Hz input vector sequence as shown in \figurename~\ref{fig:1_system_diagram}.
This corresponds to a patch size of seven, since for every $\frac{1}{12}$s, there is one token from the 12~Hz $L_0$ codebook, two from the 24~Hz $L_1$ codebook, and four from the 48~Hz $L_2$ codebook.

\subsection{Local decoder}

The global decoder's output must be converted to the full hierarchical codec tokens to vocode the output speech.
Each output vector from the global decoder is fed as the first input vector to the local decoder.
As shown in \figurename~\ref{fig:1_system_diagram}, the local decoder then autoregressively predicts each codec token for all codebooks for the current patch in a flattened way, predicting $L_0$, then two $L_1$ tokens, then the last four $L_2$ codebook tokens.

The local decoder is also a causal autoregressive transformer.
But unlike the encoder and global decoder, it always operates on a fixed sequence length of seven. 
So we use fixed, learnt positional embeddings instead of sinusoidal embeddings.

\subsection{Training}

The model is trained end-to-end with a standard cross-entropy loss to predict the next acoustic token.
Speaker embeddings are computed from the ground truth audio during training, while during inference they are computed from a desired reference speaker.
The local decoder is applied in parallel to the global decoder outputs during training and autoregressively during inference.
During training, an end-of-sequence token is appended to the acoustic tokens of the utterance, which the local encoder is trained to predict.

\section{Inference and Fine-Tuning Techniques}

\modelname{} is fast and small because most of its parameters operatore 
on only a 12~Hz sequence in the global decoder.
The shorter sequence can also improve stability.
But on its own, this new architecture does not solve the SLM-robustness problem.
Below we introduce and incorporate a `bag of tricks' for inference and fine-tuning to improve stability and performance.

\subsection{Fine-tuning setup}\label{subsec:fine-tuning}

We split model training into two parts: pretraining and fine-tuning.
Pretraining involves next-token prediction, as described earlier.
We then fine-tune the model %
using a curated high-quality subset of the training data.

For fine-tuning, 
we combine odds ratio preference optimization (ORPO)~\cite{hong2024orpo} %
and reverse inference optimization (RIO)~\cite{hu2024robust}.
First, we compute the pretraining model predictions on arbitrary text using reference waveforms from a high quality subset of the training data.
We then feed these outputs back to \modelname{} as references, 
with the transcript of the original reference, 
and predict the original reference audio in a cyclic way, as in~\cite{hu2024robust}.
We then rank the cyclic outputs based on character error rate and UTMOS~\cite{saeki2022utmos}, and select the worst performing outputs as `rejected' samples, and the corresponding ground truth audio as `chosen' samples for ORPO.
While not precisely the same as either the original ORPO (where both chosen and rejected samples come from model predictions) or RIO (where both the best and worst-performing cyclic outputs are used), we found this setup to yield the best results in preliminary experiments.

We also found that the model had a tendency to get stuck producing the same acoustic token -- this is why prior work incorporate semantic tokens in addition to acoustic tokens~\cite{baade24_interspeech}.
To remedy this, we incorporate
a flux loss to penalize repetitive generations~\cite{meng2024autoregressive}.
We adapt the flux loss used for the continuous autoregressive TTS~\cite{meng2024autoregressive} to discrete units, defining it as:
\begin{equation}
    \mathcal{L}_{\text{flux}} = \frac{\beta}{\epsilon + \text{CrossEntropy}(\hat{\mathbf{y}_t}, y_{t-1})}
\end{equation}
where $\beta$ is a scaling coefficient for the loss term, $\epsilon$ is a small offset added for numerical stability, $\hat{\mathbf{y}}_t$ is the decoder logit predictions at timestep $t$, and $y_{t-1}$ is the ground truth codebook index of the \textit{prior timestep}.
Intuitively, this penalizes the probability of the token in the prior timestep.
We apply this flux loss to $L_0$ codebook predictions, during both ORPO fine-tuning and pretraining, each with different weightings.

\subsection{Inference algorithms}\label{subsec:inference_algos}

We combine three inference methods.

\subsubsection{Repetition aware sampling (RAS)}
This approach from~\cite{chen2024valle2neuralcodec} is used on the local decoder predictions for positions corresponding to the $L_0$.
Using the notation of the original paper, we found $K=10, t_r=0.09$ to yield best results.

\subsubsection{Quality prefixing}
As mentioned in Sec.~\ref{subsec:encoder}, in training we prepend the original sample rate of the reference to the text to give the model am indication for output quality. In inference, we always set this to ``[48000]'' to maximize output quality.

\subsubsection{Top-p backoff sampling}
    SLM outputs can be made more stable by sampling with a low top-$p$ value.
    However, sometimes this can cause the model to still get stuck in a loop.
    We 
    alleviate this by using a backoff approach similar to the temperature backoff used by Whisper~\cite{whisper_radford2022robust}.
    Concretely, we sample with a top-$p$ of 0.2, and check the output length before vocoding. If the predicted audio is unrealistically short,
    we increment the top-$p$ by 0.2 and sample again.

\subsection{Shallow and deep cloning}

\modelname{} can clone from a reference in two ways -- \textit{shallow clone} and \textit{deep clone}.
The prior is where we compute the speaker embeddings from the reference audio and perform inference directly.
While simple, the speaker similarity is not optimal.
The latter is similar to the approach of VALL-E, where 
we assume knowledge of the reference transcript, and then assign a prefix to both the encoder and global decoder as the reference transcript and acoustic tokens, respectively.
This gives better prosody and speaker transfer from the reference, at the cost of inference time (longer sequence length).

\section{Experimental Setup}

\subsection{Evaluation data and baselines}

Many evaluation benchmarks do not capture the diversity of in-the-wild speech.
We therefore construct a new evaluation set on the emotive EARS dataset~\cite{richter2024ears}.
It includes emotional speech, different reading styles, free-form conversational speech, and non-verbal sounds recorded in an anechoic environment from 107 English speakers.
We select 43 speakers for the test set and 64 for the validation set.
Ignoring the non-verbal, free-form and `slow' utterances, we select half of the samples (audio and transcript) for each style, and pair each sample with another of the same speaker and style to serve as the voice cloning reference.
\modelname{} and the baseline models have, to the best of our knowledge, not seen any part of EARS.

We compare the 70M-parameter \modelname{} against three strong baseline models, all much larger: XTTSv2~\cite{casanova2024xtts_interspeech} (460M parameters), StyleTTS2~\cite{li2024styletts} (148M parameters), and MetaVoice-1B~\cite{MetaVoice2024} (1.2B parameters). 
We use the best available checkpoints and the best inference settings from each paper.

\subsection{\modelname{} implementation}

\subsubsection{Model}
We use standard 8-layer, 512-dimensional transformers for the encoder and global decoder, and a 4-layer local decoder.
For the two speaker embeddings, we use WavLM-SV~\cite{WavLMBaseSV2024} and the pretrained MS-CLAP~\cite{CLAP2023}.
We train the BPE tokenizer to a vocabulary size of 512.

\subsubsection{Training}
We train \modelname{} for 2M steps using AdamW~\cite{loshchilov2017decoupled} with a linearly decaying learning rate starting at 
$5\cdot10^{-4}$
(after a 10k step linear warmup) and ending at
$2.5\cdot10^{-5}$.
We use an AdamW $\beta$ of $(0.9,0.995)$, weight decay of $2\cdot10^{-2}$, and batch size of 96.

\subsubsection{Data}
We train \modelname{} on the following publically available datasets: LibriHeavy~\cite{kang2024libriheavy}, GLOBE~\cite{wang2024globe}, VCTK~\cite{Yamagishi2019CSTRVC}, AniSpeech~\cite{AniSpeech2024}, and CCv2~\cite{Porgali_2023_CVPR}.
We %
limit the
number
of utterances from each speaker to be at most 80k.
Together this results in a training dataset of roughly 46k hours.

\subsection{Evaluation metrics}

\subsubsection{Objective evaluation} We measure intelligibility using the word/character error rate (W/CER) between the predicted outputs on our EARS test set and the ground truth audio.
We obtain transcripts of the generated audio using the Whisper-base speech recognition model~\cite{whisper_radford2022robust}.
We objectively measure speaker cloning ability using the equal-error rate (EER) for a pretrained speaker verification system~\cite{xvector}.
The verification system produces a similarity score between pairs of utterances.
We compute these similarities on \textit{(ground truth reference, generated)} pairs and \textit{(ground truth reference, other ground truth)} pairs from the same speaker.
The former pairs are assigned a label of 0, and latter a label of 1.
Thsese can then be used to compute an EER as in~\cite{zhao2020voiceconversionchallenge2020}.
A higher EER indicates that it is harder to distinguish generated speech from ground truth examples of the reference speaker, 
up to a theoretical maximum of 50\%.
We also report an approximated mean naturalness metric using the pretrained UTMOS model~\cite{saeki2022utmos} predicting naturalness scores on a scale of 1-5.

\subsubsection{Subjective evaluation}
We perform two subjective evaluations using Amazon Mechanical Turk.
In the first, we collect a mean opinion score (MOS) on a 1-5 scale.
In the second, we collect a speaker similarity score (SIM) on a 1-4 scale following the protocol of the Voice Conversion Challenge 2020~\cite{zhao2020voiceconversionchallenge2020}.
From the EARS test set, we select 36 utterances from each baseline, the ground truth, and \modelname{} (both using shallow and deep clone).
We include trapping and calibration samples to filter out anomalous listeners, resulting in %
1326 ratings from 2340 unique listeners. %
For SIM, each evaluated utterance (from the baselines, \modelname{}, or actual ground truth audio) is paired with another random utterance from the same speaker and speaking style.
We present the listener these samples side-by-side and ask them to rate how similar the speaker sounds on a 1-4 scale similar to~\cite{zhao2020voiceconversionchallenge2020}.
After filtering anomalous listeners, we have %
1980 SIM ratings from 40 unique listeners.

\begin{table}[!t]
    \setlength{\tabcolsep}{3.2pt}
    \renewcommand{\arraystretch}{1.2}
    \centering
    \caption{
        Results measuring the intelligibility (W/CER), naturalness (UTMOS, MOS) and speaker similarity (EER, SIM) on our EARS test set. For MOS and SIM, 95\% confidence intervals are shown.
    }
    \tablecaptionsep
    \scriptsize
    \label{tab:1_headline_results}
    
    \begin{tabularx}{1.0\linewidth}{@{}
        L@{}
        r
        r
        c
        c
        S[table-format=1.2(2),
            table-figures-uncertainty=1,
            separate-uncertainty = true]
        S[table-format=1.2(2),
            table-figures-uncertainty=1,
            separate-uncertainty = true]
        @{}}
    \toprule
    Model & {WER$\ \downarrow$} & {CER$\ \downarrow$} & {EER$\ \uparrow$} & {UTMOS$\ \uparrow$} & {MOS$\ \uparrow$} & {SIM$\ \uparrow$} \\
    \midrule
    \textit{Testset topline} & 5.74 & 2.50 & {-} & 3.50 & 3.34(11) & 3.46(08) \\
    \addlinespace
    XTTSv2~\cite{xtts2} & 1.74 &  0.83 & 29.4 & 3.81 & 3.58(8) & 2.24(11) \\
    MetaVoice-1B~\cite{MetaVoice2024} & 30.70 & 27.41 & \ubold 31.2 & 3.13 & 2.84(11) & 2.47(11) \\
    StyleTTS2~\cite{li2024styletts} & \ubold 1.34 & \ubold 0.36 & 23.1 & \ubold 4.40 & \ubold 4.08(7) & 2.80(12) \\
    \modelname{} (deep) & 7.42 & 5.17 & 30.7 & 3.79 & 3.34(10) & \ubold 3.07(11) \\
    \modelname{} (shallow) & 3.96 & 2.38 & 23.1 & 3.65 & 3.44(8) & 2.24(11) \\
    {\, w/o RIO ORPO~\cite{hong2024orpo}} & 14.54 & 12.92 & 22.7 & 3.60 & {---} & {---} \\
    {\, w/o RAS~\cite{chen2024valle2neuralcodec}} & 7.31 & 5.73 & 24.0 & 3.76 & {---} & {---} \\
    {\, w/o quality prefixing} & 7.06 & 4.95 & 26.1 & 3.56 & {---} & {---} \\
    \bottomrule
    \end{tabularx}
\end{table}

\section{Results}

\subsection{Intelligibility and reference similarity}

The results on the EARS test set are given in Table~\ref{tab:1_headline_results}.
Results are mixed: for intelligibility, StyleTTS is a clear winner.
In terms of speaker similarity, \modelname{} using deep clone has the best SIM score, but in terms of EER, MetaVoice-1B is best.
For naturalness (MOS and UTMOS), StyleTTS2 again is the best.
But these results are perhaps a bit misleading, as can be seen by both XTTS, StyleTTS, and \modelname{} having better W/CER and UTMOS values than the ground truth test utterances.

While this requires further investigation, the audio samples on the demo give some insight.
Because the EARS %
is emotive, spontaneous, and diverse, it is less intelligible than pure read speech.
Models like StyleTTS2 and XTTSv2 appear to produce audio that is `de-emphasized' compared to that of the reference, particularly for highly emotive references.
Meanwhile, SLM-based models like MetaVoice and \modelname{} appear to clone the prosody of the reference more strongly at the cost of intelligibility, indicated by the higher speaker similarity metrics (especially for deep clone).
This effect is clearly heard when a whispered reference is used, where StyleTTS2 and XTTSv2 produce clean sounding outputs that are not whispered, while \modelname{} correctly produces a whispered output, even if it is slightly less intelligible (higher W/CER).
So, for highly expressive speech, lower W/CER numbers do not always correspond to outputs that are faithful to the reference utterance.

{We ablate the RAS, quality prefixing (Sec.~\ref{subsec:inference_algos}) and RIO ORPO fine-tuning (Sec.~\ref{subsec:fine-tuning}) in the last three rows of Table~\ref{tab:1_headline_results} by measuring the model's shallow clone performance i.t.o objective metrics. %
Removing any of the individual techniques degrades intelligibility.
Speaker similarity is also worse when removing RIO ORPO.
This shows that each technique is important for \modelname{}.
}

\subsection{Effect of reference style and cloning method}

\begin{figure}[!t]
\centering
\centerline{\includegraphics[width=1\linewidth]{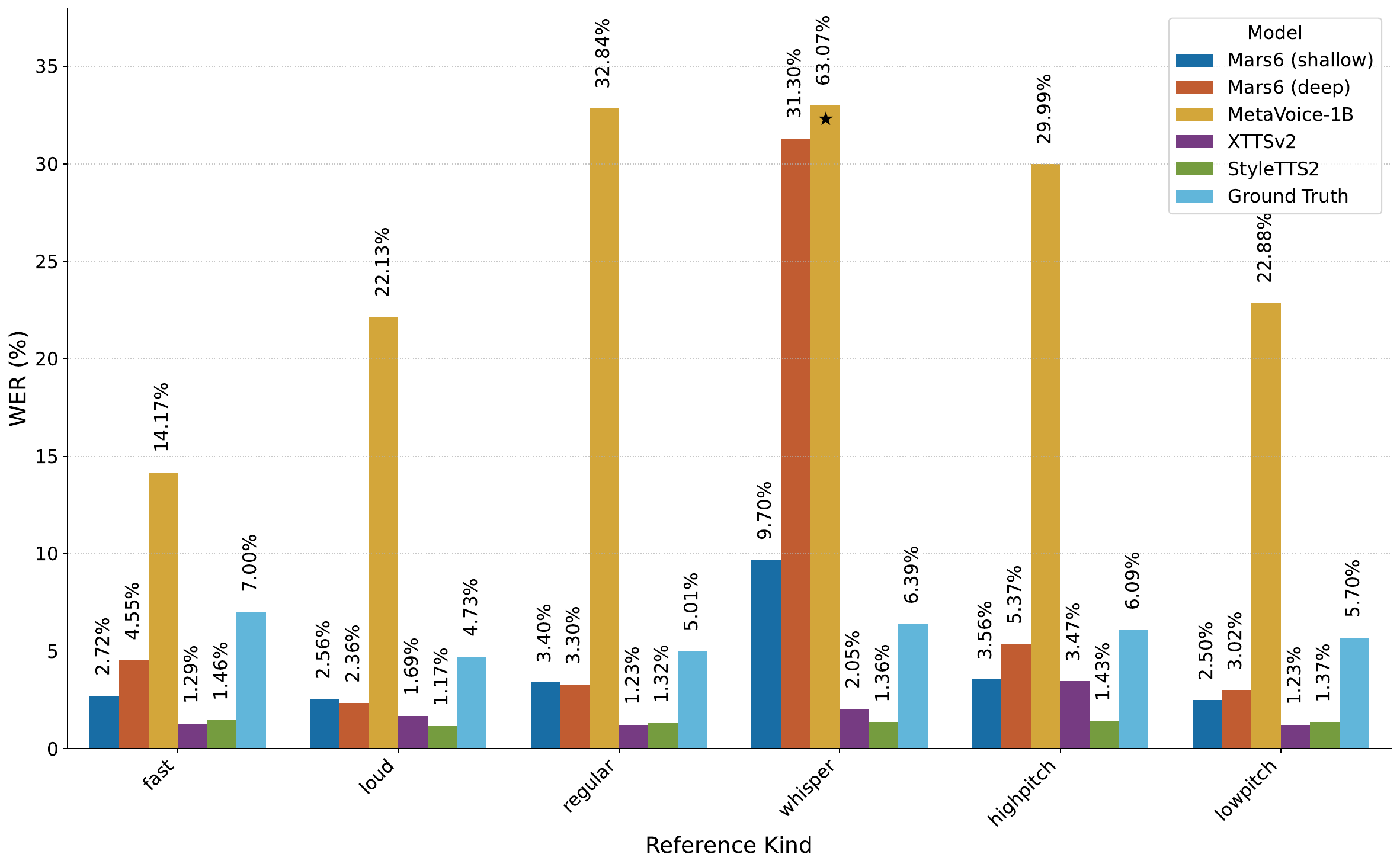}}
\vspace{-10pt}
\caption{
    Comparison of word error rates for different speaker reference styles.
}
\label{fig:2_semiblation}
\end{figure}

To demonstrate this effect a bit more, as well as profile the cases where \modelname{} is making intelligibility errors, we make use of the style labels in EARS.
Using these labels we plot the WER metric grouped by the style of the reference utterance in \figurename~\ref{fig:2_semiblation}.
The trends for most styles appear constant, except for one reference style -- whispers.
Most of the W/CER in Table~\ref{tab:1_headline_results} from both MetaVoice and \modelname{} are attributed to whispered outputs!
This, together with the audio samples, provides evidence for our earlier hypothesis.
\modelname{} is able to produce coherent whisper outputs, however, Whisper-base cannot accurately transcribe whispers.
This also causes the poorly-cloned outputs of XTTSv2 and StyleTTS2 to be rated much higher in terms of intelligibility.

\section{Conclusion}

In this work we looked to improve the robustness of discrete neural codec-based TTS models.
To this end, we proposed \modelname{}, which combines several existing and new techniques for speech language model design, training, and inference.
To evaluate robustness, we proposed a new test set built on the EARS dataset, consisting of harder and more diverse speech utterances than in other benchmarks.
We compared \modelname{} against several prior state-of-the-art TTS baselines, and found that \modelname{} achieves competitive results with models many multiples larger, particularly in terms of target speaker similarity.
Taken together, we show how many recent language and speech language modelling techniques can be effectively combined to achieve a compact, robust, and expressive TTS model.

\bibliographystyle{IEEEtran}
\balance
\bibliography{references}

\end{document}